\documentclass[aps%
 ,twocolumn%
 ,secnumarabic%
,amssymb, amsmath,nobibnotes, aps, pre, floatfix,superscriptaddress]{revtex4-1}

\usepackage{graphicx,epsfig}
\usepackage{amsmath}
\usepackage{amsfonts}
\usepackage{fancyhdr}
\usepackage{hyperref}
\usepackage{epstopdf}
\usepackage{longtable}
\usepackage{cleveref}

\begin{document}

\pagestyle{fancy}
\lhead{}
\rhead{V.Navas-Portella, \'A. Corral, and E.Vives}
\lfoot{}
\rfoot{}
%%%%%%%%%%%%%%%%%%%%%%%%%%%%%%%%%%%%%%
\title{Avalanches and force drops in displacement-driven compression of porous glasses}
\author{V\'ictor Navas-Portella}
\affiliation{Centre de Recerca Matem\`atica, Edifici C, Campus Bellaterra, E-08193 Bellaterra, Catalonia, Spain.}
\affiliation{Departament de Mat\`eria Condensada, Facultat de F\'{\i}sica, Universitat de Barcelona, Diagonal
645, 08028 Barcelona, Catalonia, Spain.}
\author{\'Alvaro Corral}
\affiliation{Centre de Recerca Matem\`atica, Edifici C, Campus Bellaterra, E-08193 Bellaterra, Catalonia, Spain.}
\affiliation{Departament de Matem\`atiques, Universitat Aut\`onoma de Barcelona,
Cerdanyola, Catalonia, Spain}
\author{Eduard Vives}
\affiliation{Departament de Mat\`eria Condensada, Facultat de F\'{\i}sica, Universitat de Barcelona, Diagonal
645, 08028 Barcelona, Catalonia, Spain.}
%\email{vicnavas1992@gmail.com} %optional
%\date{\today}
\begin{abstract}
Similarities between force-driven compression experiments of porous materials and earthquakes have been recently proposed. %This include the comparison of Gutenberg-Richter law, Omori law for aftersocks, productivity law and universal distribution of waiting times.% 
In this manuscript, we measure the acoustic emission during displacement-driven compression of a porous glass. The energy of acoustic-emission events shows that the failure process exhibits avalanche scale-invariance and therefore follows the Gutenberg-Richter law. The resulting exponents do not exhibit significant differences with respect the force-driven case.% the Gutenberg-Richter law, the modified Omori's law and the unified scaling law of waiting times.%
~Furthermore, the force exhibits an avalanche-type behaviour for which the force drops are power-law distributed and correlated with the acoustic emission events.
\end{abstract}

\maketitle

%%%%%%%%%%%%%%%%%%%%%%%%%%%%%%%%%%%%%%%%%%%%%%%%%%%%%%%%%%%%%%%%%%%%%%%%%%%
\section{Introduction}
Earthquakes constitute a complex phenomenon which has been studied for a long time due to their impact as natural disasters. From a fundamental point of view, statistical laws in seismology have attracted the attention not only of geoscientists but also of physicists and mathematicians due to their signs of scale-invariance. Recent works have found that some of these laws also manifest in materials which exhibit crackling noise: porous glasses~\cite{Salje2011, Baro2013}, minerals~\cite{Nataf2014} and wood under compression~\cite{Makinen2015}, breaking of bamboo-sticks~\cite{Tsai2016}, ethanol-dampened charcoal~\cite{Ribeiro2015}, confined-granular matter under continuous shear~\cite{Lherminier2015}, etc.  Due to the difference between time, space and energy scales, these analogies have originated an important interest in the condensed-matter-physics community. In general, the experimental results are based on the analysis of acoustic emission (AE) signals in the ultrasonic range, which are detected when these systems are mechanically perturbed.

Bar\'o et al.~\cite{Baro2013, Nataf2014} found statistical similarities between earthquakes and the AE during compression experiments of porous materials. In that case, the experiments were performed using the applied force as a driving parameter, which means that the force increases linearly in time (force-driven compression). Crackling noise during failure of porous materials has also been studied through computational models that show qualitative agreement with experimental results~\cite{Kun2013,Kun2014}. Within the context of structural phase transitions, it has been shown that avalanche scale-invariance manifests in different ways depending on the driving mechanism~\cite{Perez-Reche2008}. If the control variable for the driving is a generalized force,  disorder plays an important role leading to a dominant nucleation process and the criticality is of the order-disorder type. However, if the driving mechanism consists in the control of a generalized displacement, the critical state is reached independently of the disorder and by means of a self-organized criticality mechanism. These results were experimentally confirmed~\cite{Vives2009, Planes2013} based on the study of amplitude and energy distributions  in AE experiments of martensitic transformations. The influence of the driving mechanism has been studied in the slip events occuring in compressed microcrystals~\cite{Maass2015}. One question that still holds is whether the driving mechanism will influence or not the distributions of AE events in the case of failure under compression experiments. This question is important because when comparing with earthquakes, the natural accepted mechanism is that tectonic plates are driven at constant velocity at far enough distances from the faults~\cite{Larson1997}. Here we study the displacement-driven compression of porous glasses with the aim of answering this question.

When changing the driving mechanism from force to displacement, the first main macroscopic difference is that force fluctuates and shows drops that, as will be shown, correlate with AE events. Recently, Illa et al. have shown that the driving mechanism influences the nucleation process in martensitic transformations and these microscopic effects can lead to macroscopic changes in stress-strain curves in which force fluctuations appear \cite{Illa2015}. An exponentially-truncated power-law distribution has been found for torque drops in shear experiments of granular matter~\cite{Lherminier2015}. Serrations or force drops have also been studied in metallic single crystals~\cite{Lebyodkin1995}, metallic glasses~\cite{Antonaglia2014, Sun2010, DallaTorre2010} and in high-entropy alloys~\cite{Carroll2015}. These studies are essentially focused on the presence of criticality. Furthermore, Dalla Torre et al. studied the AE during the compression of metallic glasses and concluded that there exists a correlation between AE bursts and stress drops~\cite{DallaTorre2010}. In this work we provide a description of the distribution of force drops in displacement-driven compression experiments of porous glasses and a correlation between these force drops and the energy of the recorded AE events is identified.  

The manuscript is structured as follows: in Section~\ref{Experimental} the experimental methods as well as the sample details are described. Results are analysed in Section \ref{Results}, which is divided in three subsections: the first one (\ref{AEdata}) refers to the study of AE events, the second one~(\ref{Force avalanches}) focuses in the study of force drops and the third one  (\ref{scatter}) is devoted to the study of the relation between the energy of AE events and force drops. A brief summary and the conclusions are reported in Section~\ref{Conclusions}.

\section{Experimental Methods}
\label{Experimental}
Uni-axial compression experiments of porous glass Vycor (a mesoporous silica ceramics with $40\%$ porosity), are performed in a conventional test machine ZMART.PRO (Zwick/Roell). The cylindrical samples, with diameters $\Phi$ of $1$ mm and $2$ mm and different heights $H$ are placed between two plates that approach each other at a certain constant compression rate $\dot{z}$. We refer to such framework as displacement driven. Compression is done in the axial direction of the cylindrical samples with no lateral confinement. The force opposed by the material is measured by means of a load cell Xforce P (Zwick/Roell), with a maximal nominal force of 5 kN and output to a communication channel every $\Delta t=0.1$ s. Performing blank measurements in the same conditions as those of the experiments presented below, we have checked that force uncertainties are of the order of $10^{-2}$ N.  Simultaneous recording of AE signals is performed by using piezoelectric transducers embedded in both plates. The electric signals are pre-amplified ($60$ dB), band filtered (between 20 kHz and 2 MHz) and analysed by means of a PCI-2 acquisition system from Euro Physical Acoustics (Mistras Group) working at 40 MSPS. The AE acquisition system reads also the force measured by the conventional test machine through the communication channel. Recording of the data stops when a big failure event occurs, the sample gets destroyed and the force drops to zero. 

We prescribe that an AE avalanche or event starts at the time $t_{i}$ when the pre-amplified signal $V(t)$ crosses a fixed threshold of $23$ dB, and finishes at time $t_{i}+\Delta_{i}$ when the signal remains below threshold from $t_{i}+\Delta_{i}$ to at least $t_{i}+\Delta_{i}+200\mu$s. The energy $E_{i}$ of each signal is determined as the integral of $V^{2}(t)$ for the duration $\Delta_{i}$ of the event divided by a reference resistance of $10$ k$\Omega$.

Different experiments have been performed at room temperature for 13 different Vycor cylinders with different diameters and heights as well as different compression rates. We have checked that different cleaning protocols before the experiment do not alter the results. All the details related to experiments are  listed in Table \ref{table:samples}. 
\begingroup
\squeezetable
\begin{table}[h!]
\begin{tabular}{|l|r|r|r|}
\hline
\textbf{Sample} & \multicolumn{1}{l|}{\textbf{$\Phi$(mm)}} & \multicolumn{1}{l|}{\textbf{$H$(mm)}} & \multicolumn{1}{l|}{\textbf{$\dot{z}$ (mm/min)}} \\ \hline
V105 & 1 & 0,5 & $2\times 10^{-3}$ \\ \hline
V11 & 1 & 1 & $2\times 10^{-3}$ \\ \hline
V115 & 1 & 1,5 & $2\times 10^{-3}$ \\ \hline
V12 & 1 & 2 & $2\times 10^{-3}$ \\ \hline
V125 & 1 & 2,5 & $2\times 10^{-3}$ \\ \hline
V205 & 2 & 0,5 & $1\times 10^{-2}$ \\ \hline
V21 & 2 & 1 & $1\times 10^{-2}$ \\ \hline
V22 & 2 & 2 & $1\times 10^{-2}$ \\ \hline
V23 & 2 & 3 & $1\times 10^{-2}$ \\ \hline
V26 & 2 & 6 & $1\times 10^{-2}$ \\ \hline
V28 & 2 & 8 & $1\times 10^{-2}$ \\ \hline
V212 & 2 & 12 & $1\times 10^{-2}$ \\ \hline
V24 & 2 & 4 & $5\times 10^{-2}$ \\ \hline

\end{tabular}
\caption{Summary of dimensions and compression rates $\dot{z}$ for the different experiments reported in this work.}
\label{table:samples}
\end{table}
\endgroup
\begin{figure}
\includegraphics{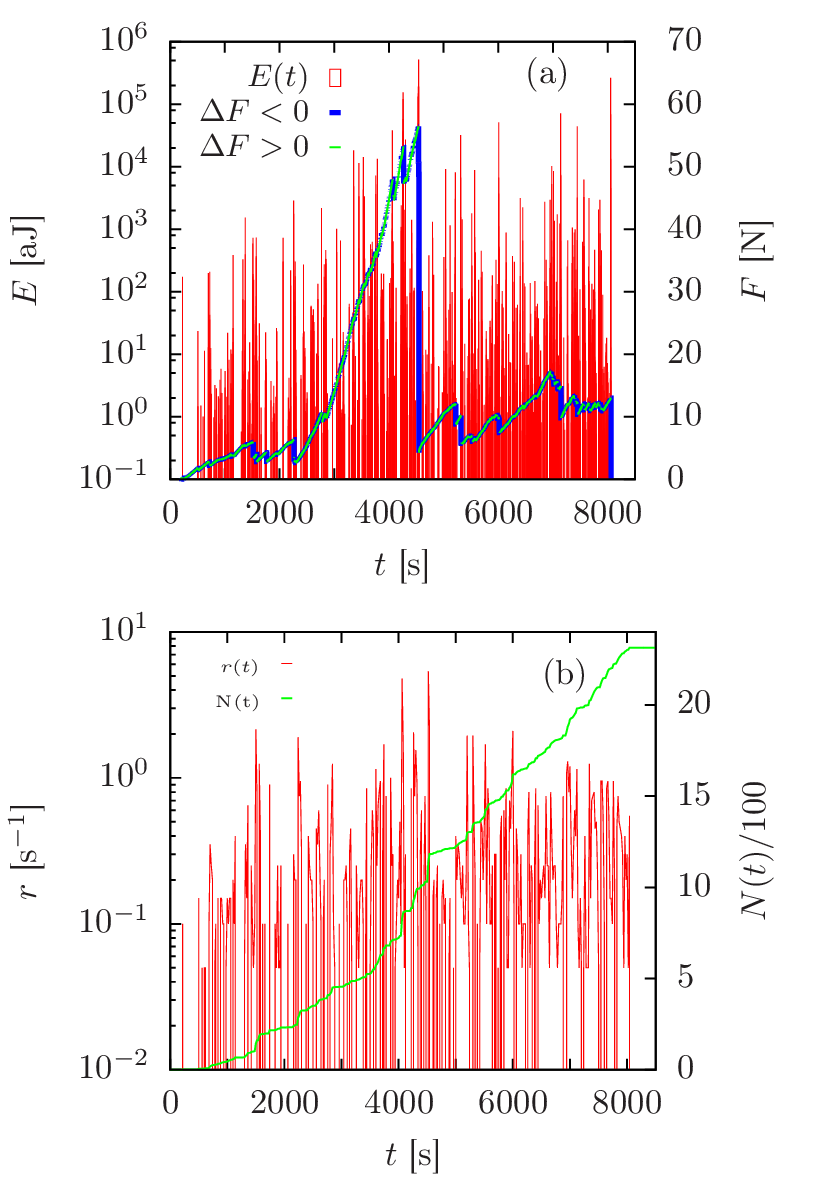}
\caption{\label{fig:experimental} Typical output for the sample V12. Panel in (a) shows the energy of the AE events as well as the measure of the force as a function of time. Green lines represent those time intervals ($\Delta t=0.1$ s) in which the force increases whereas blue lines represent those for which the force decreases (force drops). Plot in (b) represents the activity rate of the experiment as well as the cumulative number of AE events $N(t)$ as a function of time. }
\end{figure}
 Figure ~\ref{fig:experimental} shows a typical experimental output for the sample V12. Panel (a) displays the sequence of energies of the AE events and the evolution of the force as a function of the time. The acoustic activity rate $r$(s$^{-1}$) has been computed as the number of events per unite time recorded along windows of $20$ seconds. Its behaviour is shown in Figure \ref{fig:experimental}(b) together with the cumulative number of events as a function of the time. It must be noticed that force drops occur along the whole curve and clearly show variability on 3-4 orders of magnitude.  In general, the largest force drops coincide with AE events with very large energy.
\section{Results}
\label{Results}
\subsection{Acoustic Emission data}
\label{AEdata}
\begin{figure}
\includegraphics{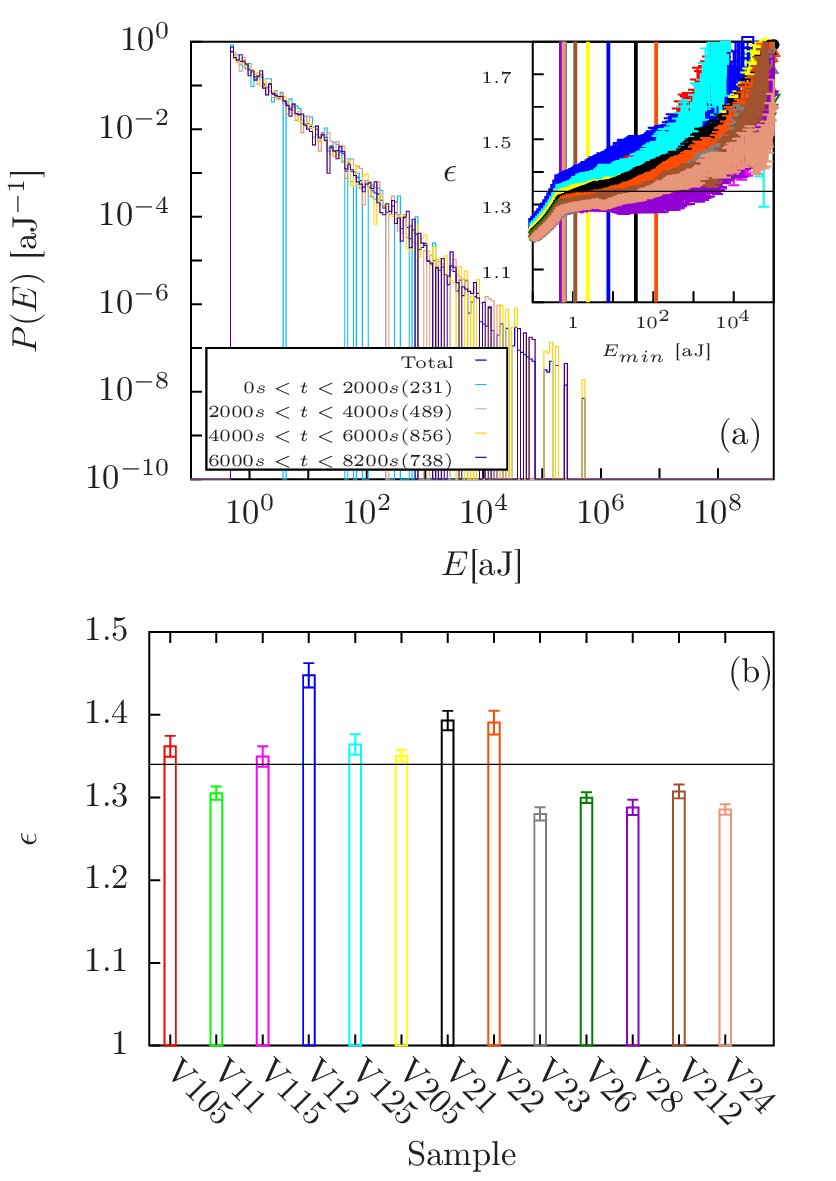}
\caption{\label{fig:energy} Panel in (a) shows the energy distribution for the sample V12 for different time windows as well as for the whole experiment. The numbers in parentheses account for the number of AE events in each time interval. Inset in (a) presents the MLE of the exponent $\epsilon$ as a function of the lower threshold $E_{min}$ for all the samples. Vertical lines correspond to the fitted values of $E_{min}$ and $\epsilon$. The color code for each sample can be read from the color bars in (b). In panel (b) the value of the exponent $\epsilon$ is shown for each sample. The dark horizontal line in the inset and in (b) is the mean value of the exponent $\epsilon=1.34$.}
\end{figure}

In force-driven compression experiments of porous glasses~\cite{Salje2011,Baro2013} it was found that the energy probability density $P(E)$ of AE events follows a power-law with exponent $\epsilon=1.39\pm 0.05$ independently of the loading rate ($0.2$ kPa/s - $12.2$ kPa/s),
\begin{equation}
P(E)dE = \left(\epsilon-1\right)E_{min}^{\epsilon-1} E^{-\epsilon} dE,
\label{eq:epower}
\end{equation}
where $E_{min} \sim 1$ aJ is the lower bound required for the normalization of the probability density. 
Figure \ref{fig:energy} (a) shows an example of histogram of the energy of AE events for the sample V12 in one of our displacement-driven experiments. As can be seen, data seems to follow the Gutenberg-Richter law for more than 6 decades. The different curves, corresponding to consecutive time windows of approximately $2000$ seconds, reveal that the energy distribution is stationary.

We use the procedure exposed in Ref.~\cite{Deluca2013} in order to guarantee statistical significance in the fit of the exponent $\epsilon$ and the lower threshold $E_{min}$.  Considering as a null hypothesis that the energy distribution follows a non-truncated power-law (see Eq.~(\ref{eq:epower})), maximum likelihood estimation (MLE) for the exponent $\epsilon$ is computed for increasing values of the lower threshold $E_{min}$ (see inset of Figure \ref{fig:energy}). For each lower threshold and its corresponding exponent, a Kolmogorov-Smirnov test of the fit is performed with a resulting $p$-value. The final values of the exponent and the threshold are chosen once the $p$-value has first overcome the significance level $p_{c}=0.05$ and the power-law hypothesis cannot be rejected. The obtained values for every sample are shown in Figure \ref{fig:energy}(b) together with the standard deviation of the MLE. The horizontal line in Figure \ref{fig:energy}(a) and in the Inset of Figure \ref{fig:energy}(a) show the average value and associated standard deviation $\epsilon=1.34\pm 0.03$. In spite of the variations around this mean value, it seems that the value of the exponent does not have a strong dependence neither on the dimensions of the sample nor on the compression rate. Complementary information obtained from the fitting method is presented in Table II.

The average value of the exponent $\epsilon=1.34\pm 0.03$ found for the present displacement-driven experiments is compatible with the value found in force-driven measurements $\epsilon=1.39\pm0.05$. Contrarily to what happens in martensitic transformations \cite{Planes2013}, we conclude that there are no clear evidences that the driving mechanism changes the value of the exponent in compression experiments. 
\begin{table}[htbp]

\begin{tabular}{|l|r|r|r|r|r|}
\hline
\textbf{Sample} & \multicolumn{1}{l|}{\textbf{$N_{AE}$}} & \multicolumn{1}{l|}{\textbf{$N_{AE}^{PL}$}} & \multicolumn{1}{l|}{$E_{min}$[aJ]} & \multicolumn{1}{l|}{$E_{Max}$ [aJ]} & \multicolumn{1}{l|}{$\epsilon$} \\ \hline
V105 & 869 & 829 & 0.602 & 1.84 $\times 10^{5}$ & 1.36 \\ \hline
V11 & 1438 & 1438 & 0.502 & 2.69$\times 10^{6}$ & 1.31 \\ \hline
V115 & 836 & 797 & 0.626 & 1.10$\times 10^{6}$ & 1.35 \\ \hline
V12 & 2314 & 928 & 7.669 & 5.16$\times 10^{5}$ & 1.45 \\ \hline
V125 & 1097 & 865 & 1.128 & 4.94$\times 10^{5}$ & 1.36 \\ \hline
V205 & 4160 & 2609 & 2.361 & 9.97$\times 10^{6}$ & 1.35 \\ \hline
V21 & 4170 & 1136 & 36.707 & 1.07$\times 10^{7}$ & 1.39 \\ \hline
V22 & 3683 & 746 & 117.583 & 6.57$\times 10^{6}$ & 1.39 \\ \hline
V23 & 1275 & 1196 & 0.645 & 7.16$\times 10^{6}$ & 1.28 \\ \hline
V26 & 2071 & 2065 & 0.516 & 1.38$\times 10^{7}$ & 1.30 \\ \hline
V28 & 974 & 974 & 0.501 & 2.82$\times 10^{6}$ & 1.29 \\ \hline
V212 & 1646 & 1338 & 1.15 & 4.37$\times 10^{6}$ & 1.31 \\ \hline
V24 & 2129 & 2039 & 0.595 & 5.97$\times 10^{6}$ & 1.29 \\ \hline
\end{tabular}
\caption{Number of AE events $N_{AE}$, number of those which are power-law distributed $N_{AE}^{PL}$, value of the lower threshold $E_{min}$, maximum value  $E_{Max}$, and exponent $\epsilon$. The standard deviation of the MLE is of the order of $10^{-2}$.}
\label{table:ae}
\end{table}

%\subsubsection{Omori Law}
%\subsubsection{Waiting times}
\subsection{Force drops}
\label{Force avalanches}
\begin{figure}
\includegraphics{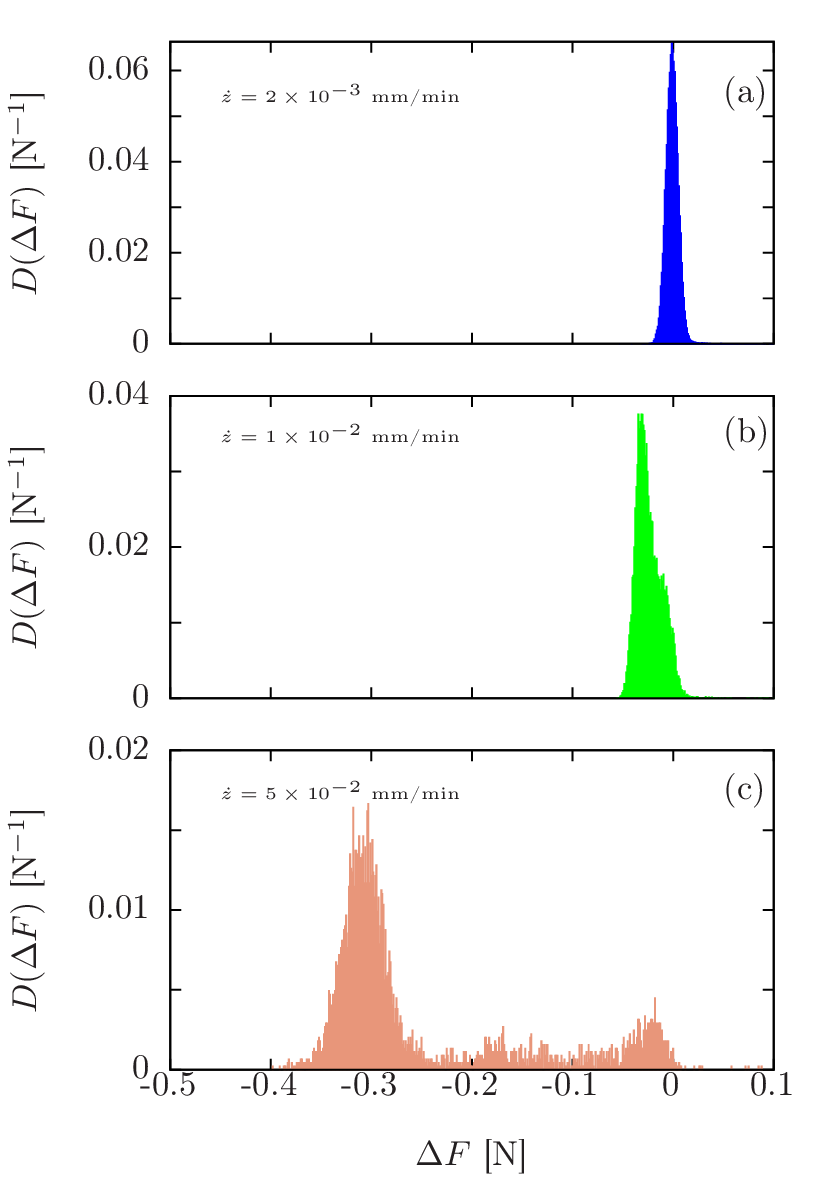}
\caption{\label{fig:global} Probability densities of $\Delta F$ for three samples with different compression rates. Sample V12 compressed at $\dot{z}=2\times10^{-3}$mm/min is shown in (a), sample V212 compressed at $\dot{z}=1\times10^{-2}$mm/min in (b) and sample V24 compressed $\dot{z}=5\times10^{-2}$mm/min is presented in (c).} 
\end{figure}

The evolution of the force as a function of time is shown in Figure \ref{fig:experimental}. We define force changes as $\Delta F(t)= -\left( F(t+\Delta t)-F(t) \right)$, with $\Delta t=0.1$ s, so that force drops are positive.  
As can be observed in Figure \ref{fig:global}(a)-(c) the distribution of $\Delta F$ can exhibit several contributions. There is a clear Gaussian-like peak corresponding to negative $\Delta F$ that shifts to the left when increasing the compression rate. This peak is related to the average elastic behaviour of the porous material. The rest of contributions in the negative part of the histogram correspond to the different elastic regimes of the material as it experiences successive failures.

\begin{figure}
\includegraphics{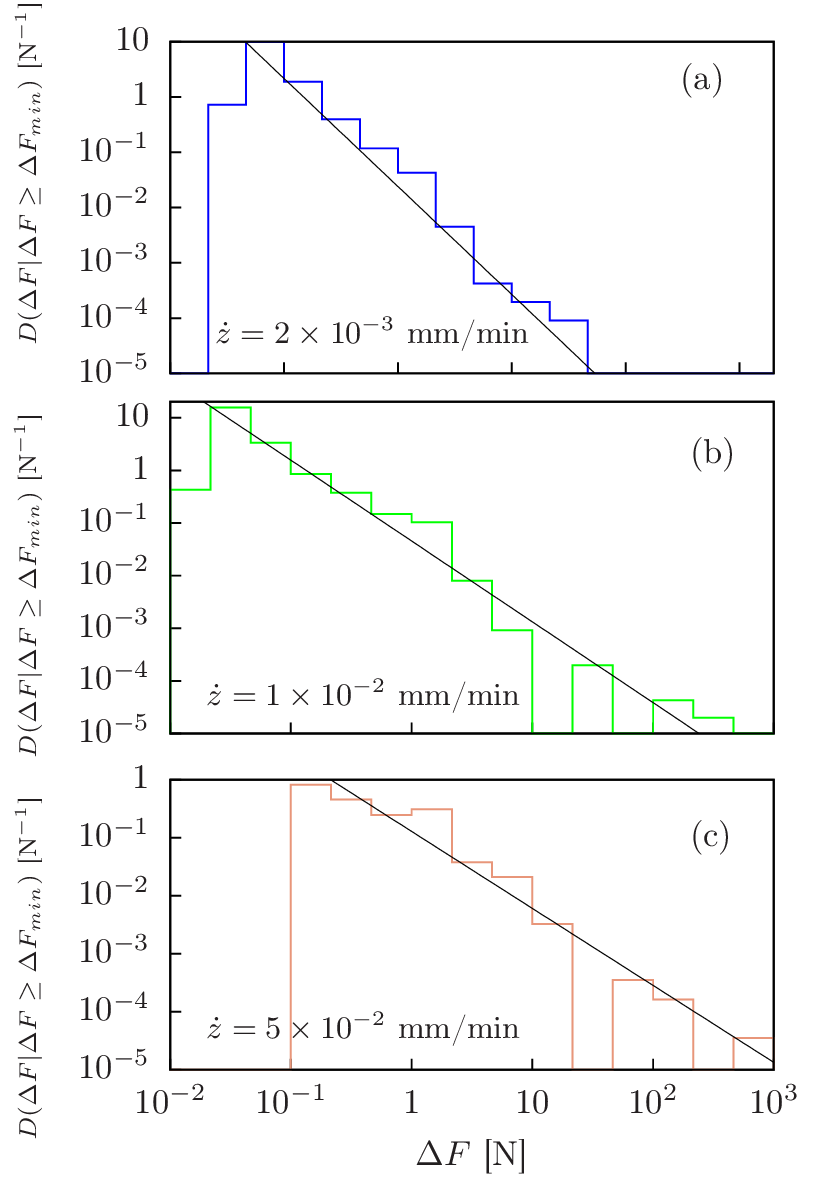}
\caption{\label{fig:dis} Probability densities of force drops $\Delta F$ and their corresponding fits for V12 (a), V212 (b) and V24 (c). Distributions are displayed and normalized for $\Delta F \geq \Delta F_{min}$.} 
\end{figure}

\begin{figure}
\includegraphics{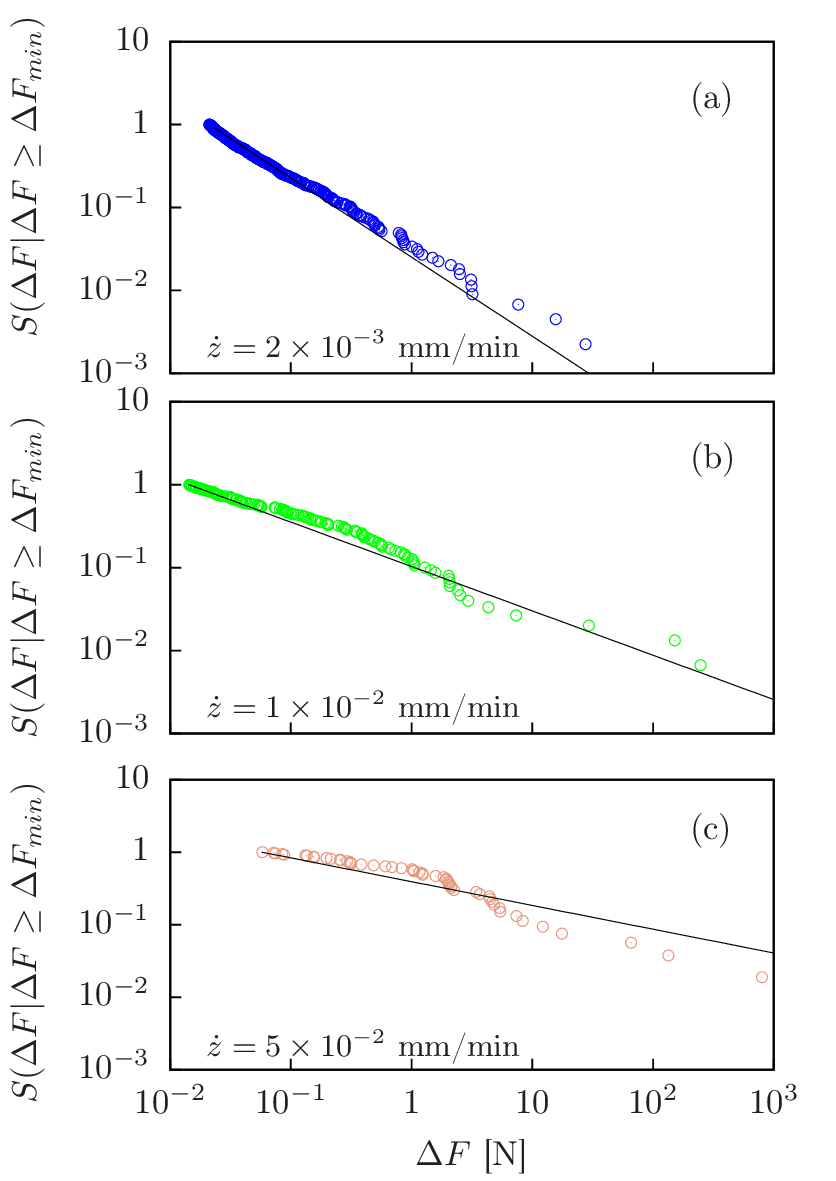}
\caption{\label{fig:deltaf} Survivor functions $S\left( \Delta F \vert \Delta F \geq \Delta F_{min} \right)$ and their corresponding fits for V12 (a), V212 (b) and V24 (c). Survivor functions are displayed and normalized for $\Delta F \geq \Delta F_{min}$.} 
\end{figure}

In the present work we will only focus on the positive part of this distribution which corresponds to the force drops. 
Our goal is to find whether the distribution of force drops is fat-tailed or not. In Figure \ref{fig:dis}(a)-(c) the distribution of force drops ($\Delta F > 0$) corresponding to Figure \ref{fig:global} is shown in log-log scale. For completeness, complementary cumulative distribution functions or survivor functions $S\left( \Delta F \right)$ are also shown in Figure \ref{fig:deltaf} (a)-(c). The probability density of force drops seems to follow a power-law $D(\Delta F) \propto \Delta F^{-\phi}$ which holds for three decades in the case of the slower compression rate and four decades for the higher ones. This difference is essentially due to the difference of surfaces of samples. The larger the surface contact between the sample and the plate, the larger the force opposed by the material.  Note that, in contrast to Fig. \ref{fig:global}, the distribution of $\Delta F$ is conditioned to $\Delta F$ larger or equal than the lower threshold $\Delta F_{min}$ obtained from the fit. 

\begin{figure}
\includegraphics{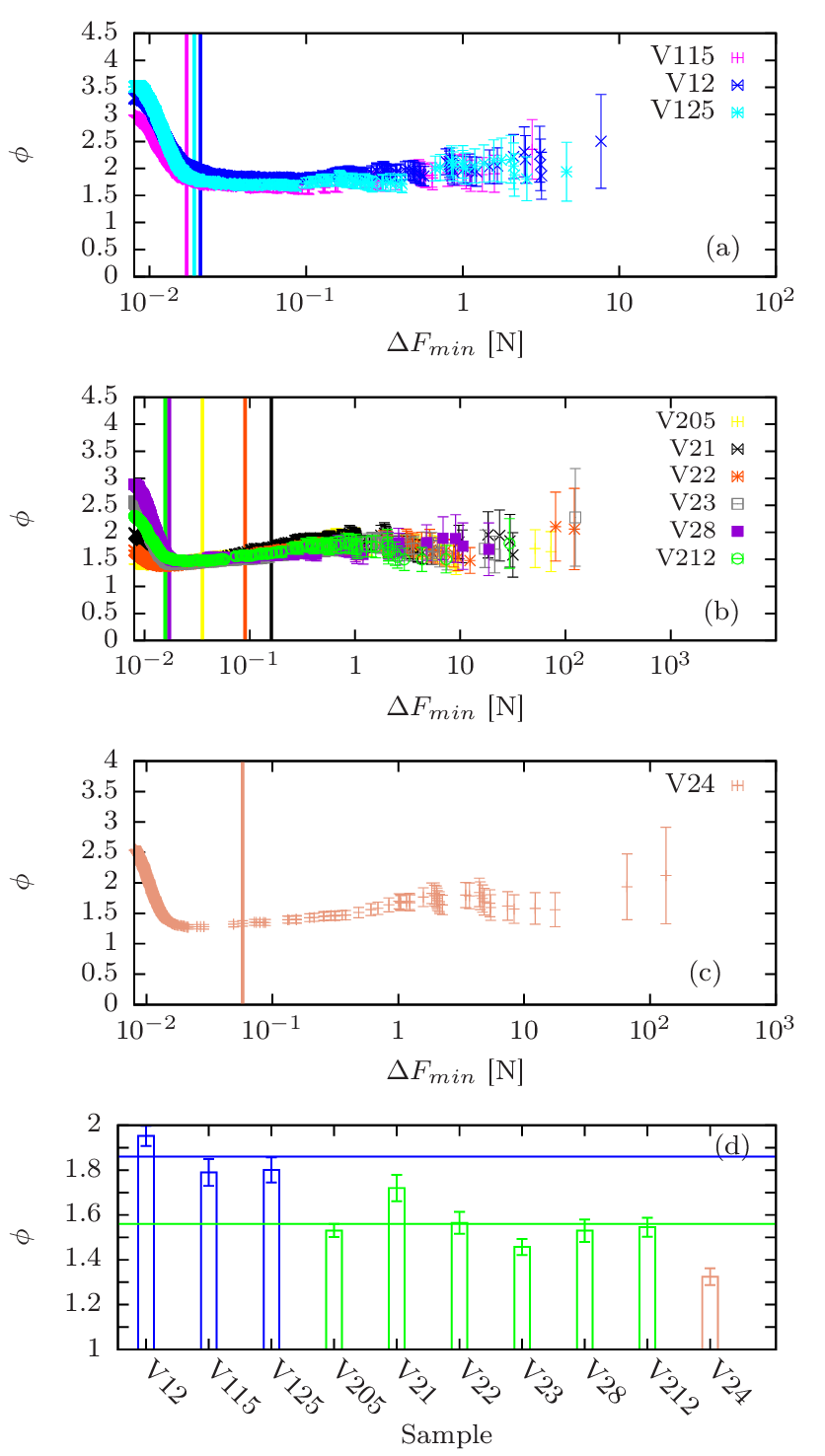}
\caption{\label{fig:fits} Panels (a)-(c) show the MLE of the exponent $\phi$ as a function of the lower threshold $\Delta F_{min}$ for samples compressed at $\dot{z}=2\times 10^{-3}$ mm/min, $\dot{z}=10^{-2}$ mm/min and $\dot{z}=5\times 10^{-2}$  mm/min respectively. Vertical lines in each panel mark the threshold $\Delta F_{min}$ which is selected by the fitting and testing procedure. Panel in (d) presents the values of the exponent for each sample. Blue horizontal line at $1.85$ and green horizontal line at $1.54$ are the mean values of the exponent for the two smallest compression rates. }
\end{figure}

In order to determine from which value $\Delta F_{min}$ the power-law hypothesis holds, the fit of the right tail of the distribution of $\Delta F$ has been performed following the same procedure as that followed for the energy distribution. %This method is much more suitable for our situation in which one does not have an easy estimation of the lower bound $\Delta F_{min}$ due to the presence of the Gaussian peaks.%
In Figure \ref{fig:fits}(a)-(c) MLE's of the exponent $\phi$ as a function of the lower threshold for the samples compressed at different compression rates are shown. Three samples have been excluded due to wrong sampling of the measurement of the force. Vertical lines of different colors represent the selected threshold $\Delta F_{min}$ for each sample. Note that, contrarily to what happens in the MLE of the energy exponent, for the lowest values of $\Delta F_{min}$ where the power-law hypothesis is not already valid, there is an overestimation of the exponent due to the presence of the Gaussian peak. 

The value of the exponents $\phi$ for the different samples is shown in Figure~\ref{fig:fits}(d) and three clear groups can be distinguished. The value of the exponent is higher for the slower compression rate and decreases for increasing compression rates. The exponent values are robust under the change of time window $\Delta t$. Additional parameters resulting from fits are shown in Table~\ref{tab:def}. %This tendency has sense since the system driven at a higher rate assembles avalanches leading to an excess of large avalanches that lowers the value of the exponent.%
\begin{table}[htbp]

\begin{tabular}{|l|r|r|r|r|r|}
\hline
\textbf{Sample} & \multicolumn{1}{l|}{\textbf{$D_{Tot}$}} & \multicolumn{1}{l|}{\textbf{$D_{PL}$}} & \multicolumn{1}{l|}{$\Delta F_{min}$[N]} & \multicolumn{1}{l|}{$\Delta F_{Max}$[N]} & \multicolumn{1}{l|}{\textbf{$\phi$}} \\ \hline
V115 & 9960 & 174 & 1.73$\times 10^{-2}$ & 8.31 & 1.79 \\ \hline
V12 & 32323 & 445 & 2.12$\times 10^{-2}$ & 27.61 & 1.95 \\ \hline
V125 & 26104 & 208 & 1.93$\times 10^{-2}$ & 24.31 & 1.80 \\ \hline
V205 & 5603 & 334 & 3.55$\times 10^{-2}$ & 853.36 & 1.53 \\ \hline
V21 & 10787 & 149 & 0.16 & 977.78 & 1.72 \\ \hline
V22 & 6609 & 133 & 9.05$\times 10^{-2}$ & 801.63 & 1.57 \\ \hline
V23 & 9987 & 162 & 1.61$\times 10^{-2}$ & 593.19 & 1.46 \\ \hline
V28 & 8881 & 113 & 1.72$\times 10^{-2}$ & 340.22 & 1.53 \\ \hline
V212 & 9030 & 202 & 1.45$\times 10^{-2}$ & 247.56 & 1.55 \\ \hline
V24 & 3742 & 53 & 5.80$\times 10^{-2}$ & 797.63 & 1.32 \\ \hline
\end{tabular}
\caption{Total number of force drops $D_{Tot}$ and the resulting values of the number of those data which are power-law distributed $D_{PL}$, values of the lower threshold $\Delta F_{min}$  and the value of the largest force drop $\Delta F_{Max}$ and the fitted exponent $\phi$. The standard deviation of the MLE is around $0.05$.}
\label{tab:def}
\end{table}

With the use of these techniques, there is evidence that force drops are power-law distributed, as found for metallic glasses \cite{Sun2010}, with a robust exponent under the change of time window and that decreases for increasing compression rates.
%\subsubsection{Waiting times between force avalanches}
\subsection{Joint distribution of Energy and Force Drops}
\label{scatter}
In this subsection we try to unveil the relation between force drops and the energy of AE events.
As it can be appreciated in Figure \ref{fig:experimental}(a), the largest force drops correspond with the highest energy of AE events. Actually, Dalla Torre et al. \cite{DallaTorre2010} found that there exists a correlation between force drops and AE events but no evidence of correlation between the amplitude of these signals and the magnitude of the force drops was found. Nevertheless, the energy could show a certain correlation since not only the amplitude plays an important role in its calculation but also the duration of the AE events.
\begin{table}[htbp]

\begin{tabular}{|l|r|r|r|r|r|r|r|}
\hline
\textbf{Sample} & \multicolumn{1}{l|}{\textbf{$U_{Tot}$}} & \multicolumn{1}{l|}{\textbf{$D_{Tot}$}} & \multicolumn{1}{l|}{\textbf{$U_{AE}$}} & \multicolumn{1}{l|}{\textbf{$D_{AE}$}} & \multicolumn{1}{l|}{\textbf{$N_{AE}$}} & \multicolumn{1}{l|}{\textbf{$N_{AE}^{U}$}} & \multicolumn{1}{l|}{\textbf{$N_{AE}^{D}$}} \\ \hline
V115 & 20119 & 9960 & 119 & 217 & 836 & 191 & 645 \\ \hline
V12 & 47663 & 32323 & 251 & 820 & 2314 & 345 & 1969 \\ \hline
V125 & 37028 & 26104 & 141 & 313 & 1097 & 215 & 882 \\ \hline
V205 & 26564 & 5603 & 1028 & 336 & 4160 & 2093 & 2067 \\ \hline
V21 & 32380 & 10787 & 1324 & 223 & 4170 & 2572 & 1598 \\ \hline
V22 & 24388 & 6609 & 930 & 177 & 3683 & 2066 & 1617 \\ \hline
V23 & 24922 & 9987 & 423 & 70 & 1275 & 804 & 471 \\ \hline
V28 & 25468 & 8881 & 359 & 83 & 974 & 638 & 336 \\ \hline
V212 & 27367 & 9030 & 453 & 135 & 1646 & 882 & 764 \\ \hline
V24 & 8114 & 3742 & 602 & 41 & 2129 & 1745 & 384 \\ \hline
\end{tabular}
\label{table:const}
\caption{Numbers which are involved in the construction of $W^{\Delta t}$. $U_{Tot}$ and $D_{Tot}$ are the total number of intervals where the force has raised up or dropped . $U_{AE}$ and $D_{AE}$ are the number of force rises and drops with AE events. $N_{AE}$ is the total number of AE events, $N_{AE}^{U}$ and $N_{AE}^{D}$ are the number of AE events associated to rises and drops of the force, respectively.}
\end{table}

\begin{figure}
\includegraphics{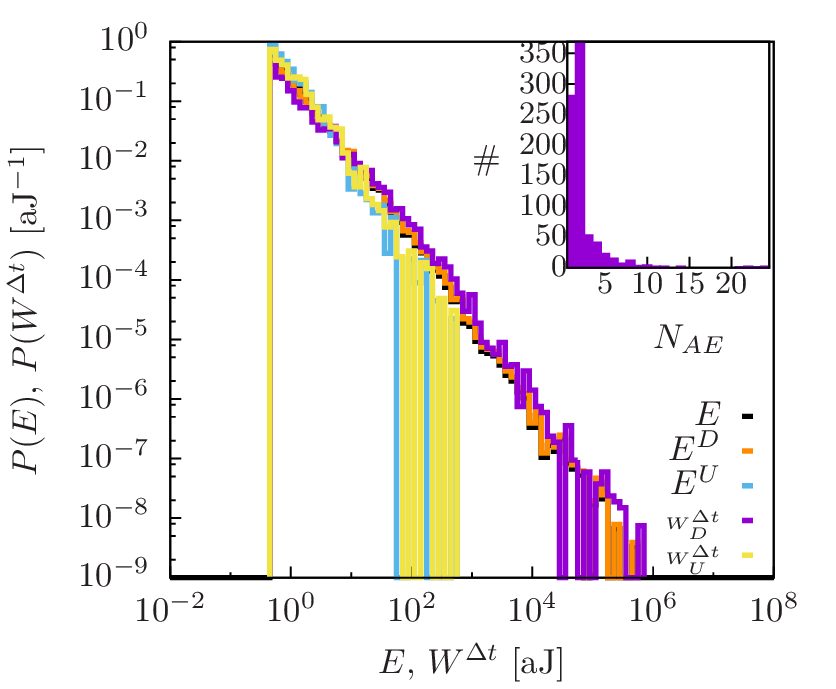}
\caption{\label{fig:marg} Main panel shows the distribution of $E$, the distributions of energies $E^{D}$ and $E^{U}$ that appear when a force drop or a force rises occurs, and the distributions of $W^{\Delta t}_{D}$ and $W^{\Delta t}_{U}$, which refer to the sum of AE energies for a certain force drop or force rise. The inset represents the histogram of the number of AE events encapsulated in time intervals where force drops occur. All these distributions correspond to the sample V12.}
\end{figure}
This correlation would be interesting for two reasons: on the one hand, it would set a relation between the energy of AE events, which is from microscopic nature (aJ), and force drops, which are at the macroscopic scale (N). On the other hand, force drops appear every time there is a micro-failure in the sample and thus they can be understood as releases of elastic energy. In the same way as Ref.~\cite{DallaTorre2010}, we find that there is a correlation in time between the occurrence of force drops and the presence of AE events. 

In order to associate a certain energy to the $i$-th force drop, we define the quantity:
\begin{equation}
W_{D,i}^{\Delta t}=\sum_{j=1}^{N_{AE}^{i}} E_{j},
\label{eq:w}
\end{equation}
where $N_{AE}^{i}$ is the number of AE events that occur within the time interval of duration $\Delta t=0.1$ s where the $i$-th force drop appears and $E_{j}$ is the energy of those AE events. The same construction can be done for force rises by defining $W_{U}^{\Delta t}$. This construction is divided in two steps: the first one consists in splitting the time axis in intervals of duration $\Delta t$ so that there is a correspondence between AE events and force rises or drops. The second step consists in applying Eq.~(\ref{eq:w}) and its counterpart for $W_{U}^{\Delta t}$ for every interval with AE events. In Figure \ref{fig:marg}(a) we present the different distributions involved in this construction for the sample V12. There are two random variables corresponding to the first step of the transformation: $E^{D}$ corresponds to the energy when a force drop appears whereas $E^{U}$ corresponds to the energy when force rises appear. The second step of the transformation is reflected in the quantities $W_{D}^{\Delta t}$ and  $W_{U}^{\Delta t}$, which correspond to the sum of energies in every force drop and in every force rise respectively. The plot in Figure \ref{fig:marg}(a) reinforces the importance of the relation between force drops and AE events since the distributions of $E^{U}$ and $W_{U}^{\Delta t}$ are restricted to low values of the energy whereas the range of the distributions of $E^{D}$ and $W_{D}^{\Delta t}$ is very similar to the original one. The inset shows the histogram of the number $N_{AE}$ of AE events encapsulated in time intervals of $\Delta t$ in which there are force drops for the sample V12. The maximum of this histogram is at $N_{AE}=2$ and decreases up to the maximum encapsulation of $N_{AE}=24$.

\begin{figure}
\includegraphics{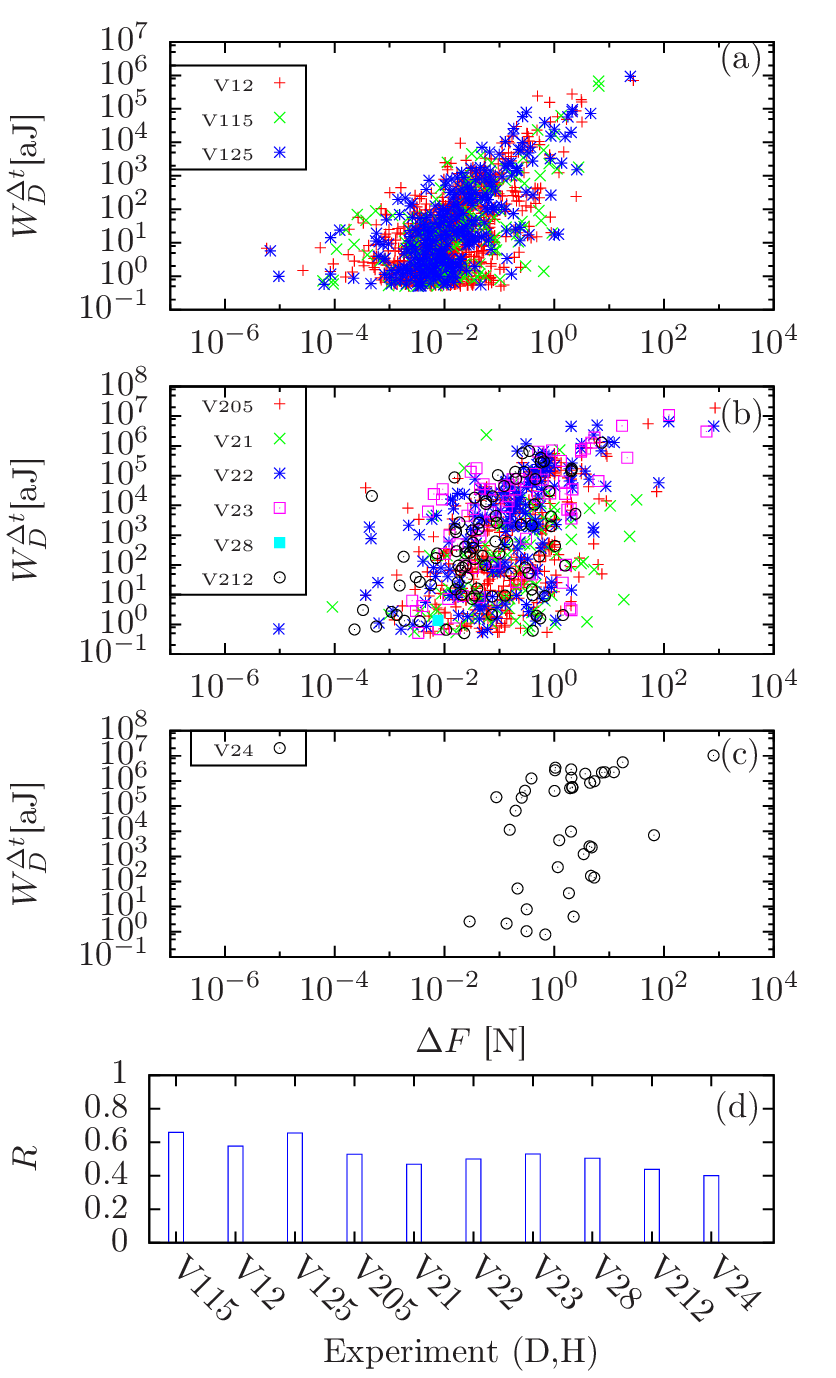}
\caption{\label{fig:scatter} Scatter plots of the energy released in each force drop for all the samples compressed at $\dot{z}= 2\times 10^{-3}$ mm/min in (a), at $\dot{z}=10^{-2}$ mm/min in (b) and at $\dot{z}=5\times 10^{-2}$ mm/min at (c). Panel in (d) shows the Pearson correlation for the logarithm of the variables for all the samples.}
\end{figure}

The numbers involved in these constructions are shown for all the samples in Table IV. The fact that there are force rises associated to acoustic emission activity can be explained by the presence of force drops that have not been identified in a $\Delta t$ interval where the force has globally increased. This prediction agrees with the fact that the energy associated to force rises covers a small range corresponding to low energy values of the total energy distribution. It is important to remark that, despite the fraction of AE events associated to force drops decreases as the compression rate increases, the fraction that accounts for the average number of events encapsulated in a force drop ($N_{AE}^{D}/D_{AE}$) is always larger than the average number of AE events encapsulated in intervals where the force is increasing ($N_{AE}^{U}/U_{AE}$). Hence, increments of AE activity are essentially associated to drops in the force. The total duration of the experiment is given by $T=\left( U_{Tot}+D_{Tot}\right)\Delta t$. Note that, despite the big difference between the total number of force drops ($D_{Tot}$) and the number of force drops with AE activity ($D_{AE}$), this second number is in the same order of magnitude as the number of power-law data in Table~\ref{tab:def} but larger always.
In Figure \ref{fig:scatter} we present scatter plots for the different compression rates. 
It must be noticed that the largest AE events are manifested in those force drops which are power-law distributed. The associated energy of the remaining force drops is relatively low compared with those with large values of $\Delta F$. The rest of force drops that have no associated AE activity  are related to experimental fluctuations of the measurement.
Under these circumstances, we study the energy associated to force drops and try to unveil if there exists any correlation between them. It must be mentioned that, as it has been seen in the previous section, the range of interest of force drops is restricted to those values which exceed $10^{-2}$ N.
In Figure \ref{fig:scatter}(d) the Pearson correlation of the logarithm of the variables for the range of interest is shown for each sample. %It has been proved through reshuffling of data that these correlations are not numerical artefacts.% 
These correlations are much higher than the ones resulting after the reshuffling of the data, so they have statistical significance. The correlation is positive and it establishes a relation between AE events, which are of microscopic nature, with a magnitude of macroscopic character, the force drops.

\section{Conclusions}
\label{Conclusions}
In this manuscript we have reported the results of displacement-driven compression experiments of several Vycor cylinders with different dimensions and different compression rates. The Gutenberg-Richter law is found for the energy distribution in the same way it was previously found for force-driven compression experiments. Regarding the values of the exponents, we conclude that they do not seem to be affected by the driving mechanism in compression experiments. The independence with the driving mechanism  has also been found in the measurement of slip events in microcrystals \cite{Maass2015}.

When the driving variable turns out to be the displacement, the release of elastic energy is not only expressed by means of AE but it is also manifested as drops in the force which are power-law distributed with a compression-rate-dependent exponent. These drops can also be observed in computer simulations near the big failure event \cite{Kun2013, Kun2014}. Nevertheless, some tuning of the disorder should be arranged in simulations in order to replicate a situation with a similar level of heterogeneity as in our experiments. Furthermore, we have established a correlation between force drops and the associated energy of AE events.
\begin{acknowledgments}
We  thank Jordi Bar\'o and Ferenc Kun for fruitful discussions.
The research leading to these results has received funding from "La Caixa" Foundation. Financial support was received from FIS2012-31324, FIS2015-71851-P, MAT2013-40590-P, Proyecto Redes de Excelencia 2015 MAT2015-69-777-REDT from Ministerio de Econom\'ia y Competitividad (Spain) and 2014SGR-1307 from AGAUR.   
\end{acknowledgments}

%\bibliographystyle{AIP}
%\nocite{*}
\bibliography{Displacement}

\end{document}